\documentclass[conference]{IEEEtran}
\IEEEoverridecommandlockouts
\usepackage[pdftex]{graphicx}
\usepackage{multirow}
\usepackage{amsmath}
\usepackage{siunitx}
\usepackage{color}
\usepackage{url}

\newcommand{\eg}{\textit{e.g.}}
\newcommand{\ie}{\textit{i.e.}}
\newcommand{\etal}{\textit{et al.}}

\newcolumntype{L}[1]{>{\raggedright\let\newline\\\arraybackslash\hspace{0pt}}m{#1}}

\begin{document}

\title{Intrusion Detection at Scale with the Assistance of a Command-line Language Model
}

\author{
Jiongliang Lin\textsuperscript{\rm 1}\quad Yiwen Guo$^\dagger$\thanks{$^\dagger$Yiwen Guo is the corresponding author.}\quad Hao Chen\textsuperscript{\rm 2}\\
{\textsuperscript{\rm 1}Sun Yat-sen University, China \  \textsuperscript{\rm 2}UC Davis, US}\\
\tt{\small{linjliang3@mail2.sysu.edu.cn, guoyiwen89@gmail.com, chen@ucdavis.edu}}
}

\maketitle

\begin{abstract}
Intrusion detection is a long standing and crucial problem in security. A system capable of detecting intrusions automatically is on great demand in enterprise security solutions. Existing solutions rely heavily on hand-crafted rules designed by security operators, which suffer from high false negative rates and poor generalization ability to new, zero-day attacks at scale. AI and machine learning offer promising solutions to address the issues, by inspecting abnormal user behaviors intelligently and automatically from data. However, existing learning-based intrusion detection systems in the literature are mostly designed for small data, and they lack the ability to leverage the power of big data in cloud environments. In this paper, we target at this problem and introduce an intrusion detection system which incorporates large-scale pre-training, so as to train a large language model based on tens of millions of command lines for AI-based intrusion detection. Experiments performed on 30 million training samples and 10 million test samples verify the effectiveness of our solution.
\end{abstract}

\begin{IEEEkeywords}
Intrusion detection, anomaly detection, language model, large-scale pre-training, command lines
\end{IEEEkeywords}

\section{Introduction}

\begin{figure}[th]
    \centering
    \vskip -0.1in
    \includegraphics[width=0.94\linewidth]{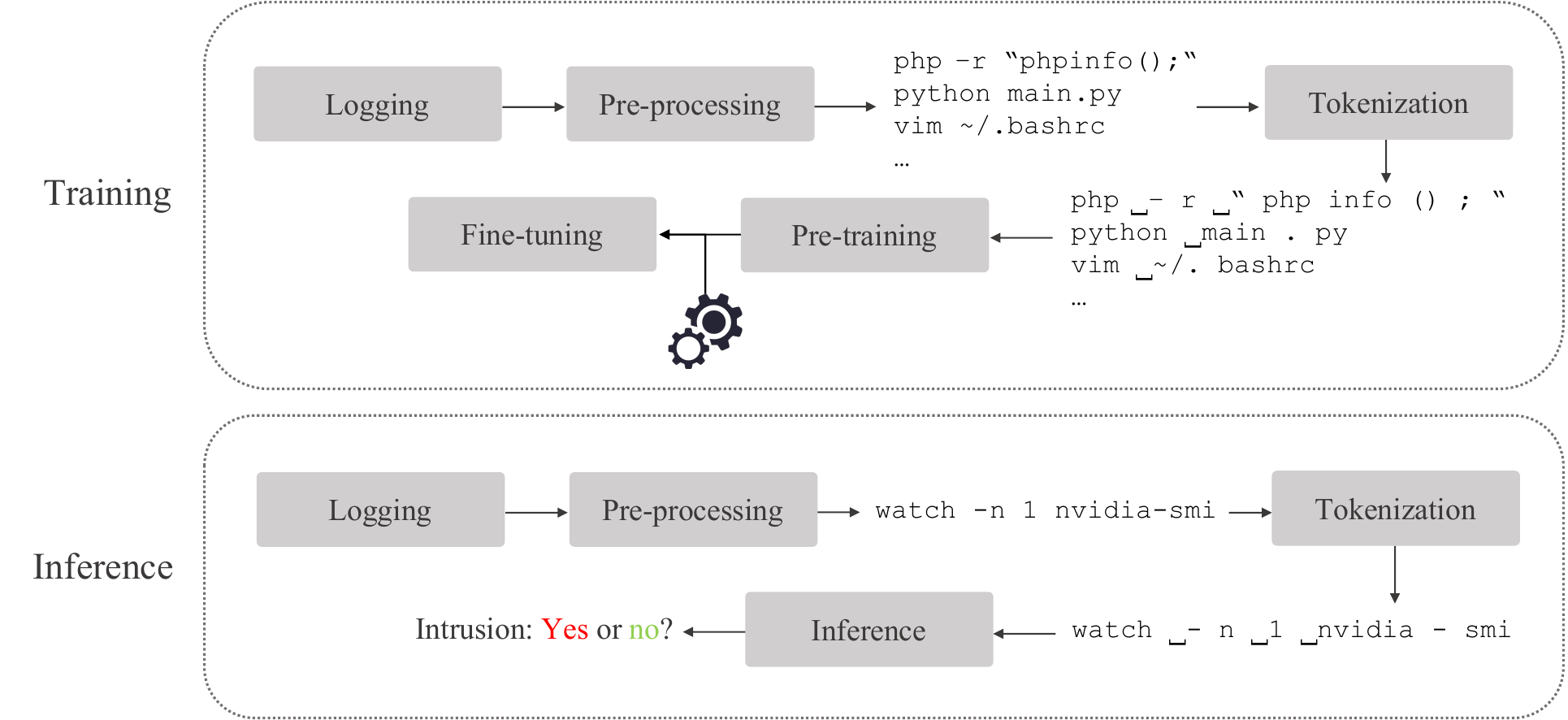} \vskip -0.1in
    \caption{The training and inference pipeline of our IDS.} 
    \label{fig:pipeline} 
\end{figure}

In this paper, we introduce an intrusion detection system (IDS) that incorporates large-scale self-supervised pre-training to train a language model that could release the power of big data for intrusion detection. We take advantage of user command lines to train a language model. 
Equipped with the command-line language model, we are capable of building an IDS to continuously learn from tens of millions of user command lines every week for digging out future attacks and intrusions. Figure~\ref{fig:pipeline} illustrates the pipeline.

To verify its effectiveness, we performed an experiment on \textbf{more than 30 million} training command lines collected in a production cloud environment, and evaluated the performance of our solution on other 10 million command lines. To the best of our knowledge, this is the first attempt for adopting advanced AI models to such large-scale data for intrusion detection in the literature. Systematic study has been performed, and experimental results show the effectiveness of our IDS at scale. Specifically, tuning the model leads to \textbf{a prediction precision of $\mathbf{>83\%}$ in digging out intrusions that should have been detected (yet missed unfortunately) by a commercial IDS}, while ensuring that intrusions previously confirmed by the commercial IDS are also spotted. The overall prediction precision of the method is $\mathbf{>99\%}$, making it possible to be used as a competitive replacement of the commercial IDS. 
Qualitative analyses also show the effectiveness of the developed methods. 
Intriguing examples are provided to shed light on how the model helps. 

The key contributions of this paper are threefold. First, we marry self-supervised learning with intrusion detection, which are core problems in different communities. Second, we, probably for the first time, develop a language model to specifically understand event logs and command lines at scale. Third, methods have been proposed to adapt the model to intrusion detection in practice, leading to superior performance in comparison with a commercial IDS in a systematic and large-scale evaluation.

\section{A Language Model for Event Logs}
\label{sec:method}

There exist an extremely large amount of unlabeled data in cloud environments, and we consider introducing self-supervised pre-training to IDS, to fully release the power of such big data.
At present, cloud computing solutions employ a command-line interface that receives text-based commands from users. 
All terminal behaviors of a user, either benign or adversarial, can be fully recorded by logging command-line operations, and intrusion detection can further be performed by analyzing his/her command-line operations. 
Hence, we focus on command lines in this work. The solutions based on command lines can be combined with IDSes that analyze other signs of adversarial activitivies (\eg, the existence of malwares~\cite{christodorescu2005semantics, ye2017survey}), and it can be studied in future work. 

\subsection{Data Logging and Pre-processing}\label{sec:preprocess}

\begin{figure}[ht]
    \centering
    \vskip -0.1in
    \includegraphics[width=0.92\linewidth]{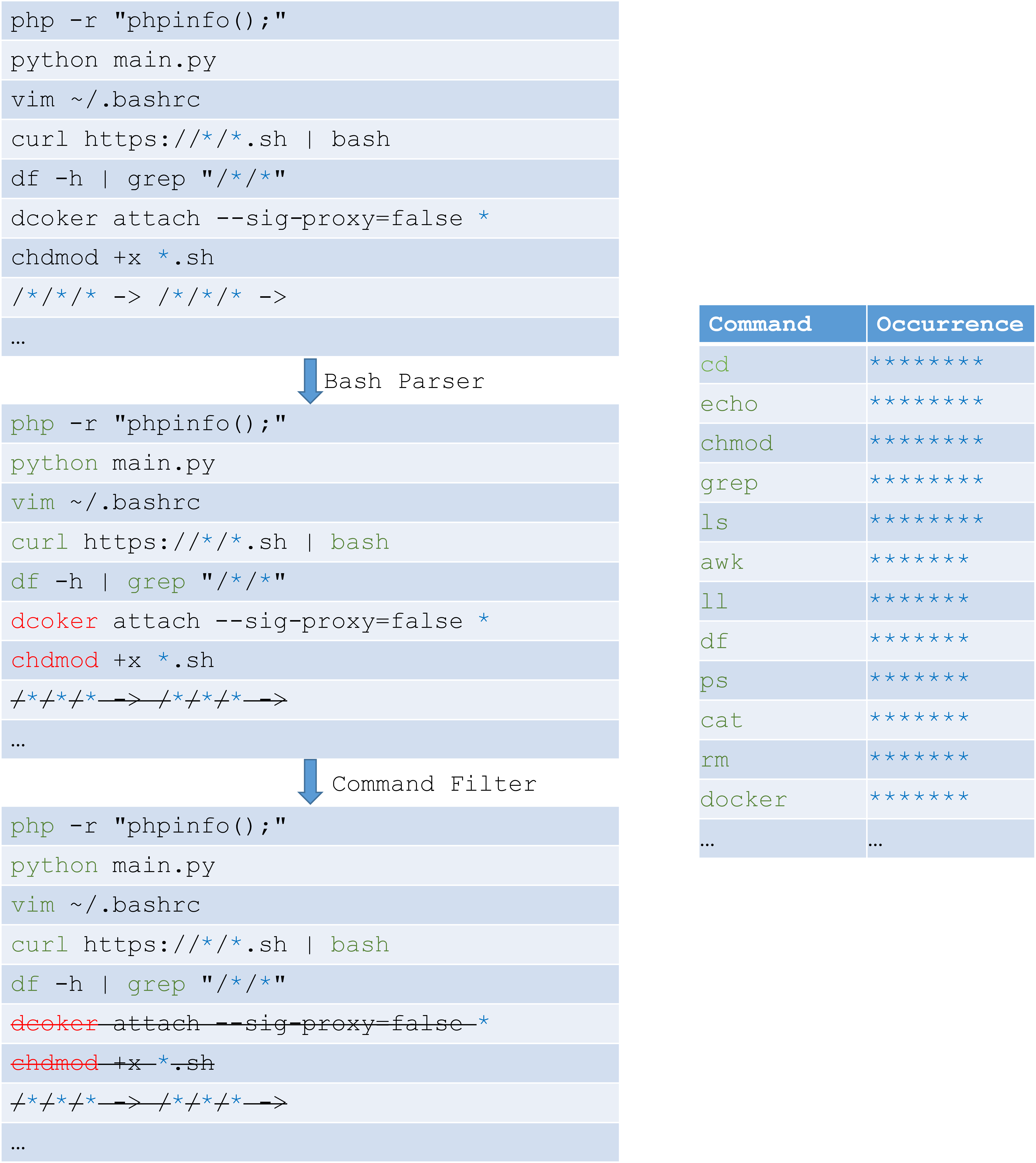} \vskip -0.1in
    \caption{Pre-processing using a parser and a list of concerned commands. Invalid command lines will be removed from further processing.} \vskip -0.1in
    \label{fig:preprocess}   
\end{figure}

Data should be collected in the first place for training language models (see Figure~\ref{fig:preprocess}).
Since typos inevitably exist, a non-negligible portion of the command-line data cannot be successfully executed by the system, and therefore can hardly be harmful to the system. Removing such command-line records from deeper analyses not only improves run-time performance of the IDS, but can also be beneficial to training AI models required by the IDS. 
To achieve this, one can simply utilize a parser to try parsing each command line into unified types. 
A parser converts each command line into a tree structure with command nodes, and it is capable of separating command names from flags and arguments. As shown in Figure~\ref{fig:preprocess}, command lines like ``\texttt{/{\color{blue}*}/{\color{blue}*}/{\color{blue}*} -> /{\color{blue}*}/{\color{blue}*}/{\color{blue}*} ->}'' can thus be removed, since it involves an invalid redirection operator ``\texttt{->}''. Nevertheless, it can also be observed in the figure that a parser may incorrectly treat ``\texttt{dcoker}'' as a command name, which is in fact a typo for ``\texttt{docker}'' (though ``\texttt{docker}'' should appear much more frequently than ``\texttt{dcoker}'' in logs). In order to address this, one can introduce a list of concerned command names, either by exhaustively collecting all valid commands in the host environment or by filtering out data that shows extremely low frequency and thus is less likely to be valid. Figure~\ref{fig:preprocess} illustrates the pre-processing steps.

\subsection{Large-scale Self-supervised Pre-training}
\label{sec:pretrain}

Just like in natural language processing, where identification of contiguous characters that form a semantic unit (\eg, words and subwords called tokens) in each sentence is required, here we also need to divide command lines of their original string format into smaller units (\ie, tokens) that somehow align with the Bash command syntax. In this work, we use the BPE tokenizer~\cite{sennrich2015neural}.

After tokenization, each command line is separated into a sequence of sub-strings, \ie, tokens. Each token can be represented by a unique one-hot vector indicating its token ID. 
Training on such data at scale in a self-supervised manner has been studied, which has inspired mask language modeling (MLM)~\cite{devlin2018bert,liu2019roberta} and causal language modeling~\cite{brown2020language}. Since MLM excels in downstream discriminative tasks, we focus on this line of work and attempt to adapt them to command line data. 
We use the same masking strategy as in RoBERTa~\cite{liu2019roberta}. That is, at each pre-training iteration, each token in the training command lines will be replaced with a ``\texttt{[MASK]}'' token, in a probability of $q$. The learning model is encouraged to reconstruct the randomly masked token using those unmasked ones which are generally informative enough to accomplish the task. Imagining that when \texttt{[MASK] http://{\color{blue}*}/{\color{blue}*}.sh \textbar{} bash} is presented, practitioners are easy to infer that this command line is probably crafted for downloading a bash script and execute it, and those familiar with the command-line interface should know that the masked token is likely to be \texttt{curl} or \texttt{wget}. This process does not require any human annotation and is considered as self-supervised.

The transformer architecture~\cite{vaswani2017attention} is chosen for training, which takes a sequence of summations of the token encoding and positional encoding vectors as input and it produces a sequence of feature representations as output. Each of the feature representations, which is a vector, will be referred to as a token embedding in this paper.

(Pre-)training of the transformer and encoding vectors, at each iteration, often takes a batch of samples together as model inputs, and an average of the MLM loss over all these samples is calculated. The model is trained to minimize such average loss. Backpropagation~\cite{rumelhart1985learning,lecun1988theoretical} and stochastic gradient descent~\cite{bottou1998online} are utilized for computing gradients and updating parameters in $\theta$.

\section{Unsupervised Anomaly Detection Using the Command-line Language Model}
\label{sec:anomaly_dec}

The model pre-trained as in Section~\ref{sec:pretrain} can be regarded as a powerful encoder~\cite{liu2019roberta}, and we can easily obtain semantic embedding of any command line from the output of such an encoder. 

There exist several different ways of utilizing such a model and the produced embeddings. 
One option is unsupervised anomaly detection~\cite{vaswani2017attention}.
The core assumption for such a line of methods to succeed is the rare occurrence of anomaly (which should be malicious command-line operations related to intrusions in our scenario). Typical unsupervised anomaly detection methods, including one-class support vector machines~\cite{scholkopf2001estimating}, isolation forest~\cite{liu2008isolation}, and principal component analysis (PCA)~\cite{wall2003singular}, can be performed in the embedding space directly or in some modified embedding spaces. 

Taking PCA as an example, when command-line embeddings are utilized, PCA can be applied to detect anomalies by evaluating the reconstruction error defined as follows:
\begin{equation}\label{eq:recons}
L_{\mathrm{PCA}}(\mathbf t) = \|W^TWf(\mathbf t)-f(\mathbf t)\|^2_2,
\end{equation}
where $W$ is a $p\times q$ projection matrix which could map the $q$-dimensional command-line embedding $f(\mathbf t)$ to a more compact space of dimensionality $q>p$. There exist several different ways of obtaining the command-line embedding $f(\mathbf t)$~\cite{reimers2019sentence, li2020sentence}, and one can simply perform average pooling to aggregate information in all token embeddings of the command line. The PCA projection matrix $W$ can be easily obtained via singular value decomposition (SVD)~\cite{wall2003singular}. 

Certain intrusions can be successfully detected using such an unsupervised method. For instance, ``\texttt{masscan {\color{blue}*} -p 0-65535}'' that scans all ports associated with an IP address (which is replaced with a ``\texttt{\color{blue}*}'' symbol for anonymity reasons) shows a very high reconstruction error of roughly $230$. With such a high reconstruction error, it emerges in the top-10 highest rated command lines among 10 million test samples. Nevertheless, we also observed that a non-negligible set of the test samples show high reconstruction error only on account of their complex arguments or inferior user habits of command-line interactions, \eg, the ``\texttt{mv}'' command followed by a very large number of complex filenames and the ``\texttt{echo}'' command followed by long and weird (yet benign) texts, \eg, ``\texttt{a...ab...bc...}''. These command lines are somehow not common, in comparison with normal user behaviors that only move a very limited number of files at a time and echo human-understandable texts, yet they are NOT related to intrusions or attacks to the best of our knowledge and should be classified into benign behaviors.

\section{A Bit Supervision if Possible}\label{sec:supervision}

As has been discussed, despite its merit in simplicity, unsupervised anomaly detection sometimes suffers from gaps between ``abnormal yet benign'' human behaviors and manually defined intrusions. That is, command-line operations detected by AI models without any sort of supervision may inevitably include a large number of abnormal yet benign user operations. To address this, one can consider ways of incorporating some amount of supervision and knowledge, if available, to help AI better identify what are adverse behaviors and intrusions that should be paid attention to. 

Supervision may come from a variety of sources, though all very noisy~\cite{natarajan2013learning} in general, \eg, alerts from commercial IDSes and alerts triggered by off-the-shelf hand-crafted rules proposed by professionals. 
Such noisy supervision can be provided in a black-box manner by querying the commercial IDSes and writing regular expression, respectively, just for labeling a number of command lines.

Given a set of labeled data $\{(\mathbf t_i, y_i)\}$, in which $y_i$ being $1$ or $0$ indicates the command line $\mathbf t_i$ being labeled to be related to an intrusion or not, we consider several different ways of incorporating supervision, as follows. 

\subsection{Reconstruction-based tuning} \label{sec:recon}
First, we can follow the formulation of PCA-based unsupervised anomaly detection to compute $W$ and reconstruction errors. Then we tune learnable parameters in $f(\cdot)$ to encourage intrusion-related command lines in the labeled dataset to show high reconstruction errors while the others show low errors, \ie, $f(\cdot)$ is optimized to minimize: 
\begin{equation}\label{eq:loss_recons}
L_{\mathrm{Recons}} = -\log\frac{\sum_{i} L_{\mathrm{PCA}}(\mathbf t_i) y_i }{\sum_{i} L_{\mathrm{PCA}}(\mathbf t_i)}.
\end{equation}
By minimizing the loss in Eq.~\eqref{eq:loss_recons}, the reconstruction error (\ie, $L_{\mathrm{PCA}}$) of intrusion-related command lines is explicitly encouraged to be large, and the reconstruction error of the other samples is constrained to be relatively small. The optimization of $f(\cdot)$ using Eq.~\eqref{eq:loss_recons} and a fixed $W$ can be accomplished via backpropagation and stochastic gradient descent, just as in the pre-training phase.
Once the optimization of $f(\cdot)$ converges, we obtain a new $f(\cdot)$ function. 
We then re-compute the projection matrix $W$ using updated command-line embeddings produced by the new $f(\cdot)$, and, afterwards, the parameters of $f(\cdot)$ can further be updated via minimizing Eq.~\eqref{eq:loss_recons} with the new $W$. Such an alternative optimization process of updating $f(\cdot)$ and $W$ is performed back and forth until final convergence is achieved. This method is called \emph{reconstruction-based tuning} in our paper, since its training objective contains the reconstruction errors. 

\subsection{Classification-based tuning} \label{sec:classif}
As our second choice for incorporating supervision, we adopt probing~\cite{tenney2019bert}, which places a shallow classification head on top of the ``\texttt{[CLS]}'' embedding produced by the pre-trained command-line language model. Given the annotated dataset $\{(\mathbf t_i, y_i)\}$, we tune the head for classifying command lines correctly, \ie, to encourage the output of the classification head to be aligned with $y_i$, while keeping the backbone frozen. This method is \emph{classification-based tuning}. The optimization objective for achieving the goal is:
\begin{equation}
    L_{\mathrm{Classif}} = -\sum_{i} \log{p(y_i|\mathbf t_i;\theta)},
\end{equation}
where $p(\cdot|\cdot;\theta)$ indicates the likelihood of an input command line being classified into be intrusion-related or not. 

\subsection{Multi-line Classification} It is sometimes insufficient to spot malicious behaviors only by investigating a single command line. For instance, we can easily tell that ``\texttt{wget -c http://{\color{blue}*}/{\color{blue}*} -o python}'' followed by  ``\texttt{python}'', should be paid attention to, because the two command lines download a suspect file from the internet and rename it to ``\texttt{python}'' before executing. By contrast, if analyzing them one by one, both command lines may seem less malicious, and it will become more difficult to judge even for practitioners. On this point, we further try \textit{multi-line classification-based tuning}. More specifically, for classifying a particular command-line operation, several command lines in the most recent past from the same user are additionally served for reference, if their execution time is not too long ago. These command lines are concatenated with a shell command separator ``\texttt{;}'' before being fed into the model, and the model is then tuned similarly with supervision. 

\subsection{Retrieval-based method}
In addition to these tuning-based methods, we further consider retrieval-based detection to utilize the supervision and the command-line language model without any tuning. Generally, the $k$-nearest neighbors algorithm ($k$NN)~\cite{cover1967nearest} shall be the first to be tried. Indeed, we can perform $k$NN in the embedding space of the model. Since it is a non-parametric method, it demands no tuning of the pre-trained model. In particular, for each test command line, if the majority of its $k$ nearest neighbors in the training set have been annotated as malicious by the supervision source, then one may consider the test sample to be malicious as well. By contrast, if not, then it is treated as benign by the method. The prediction score of such a method is the average similarity of those (known to be) malicious neighbors and the test sample. 
Although such a method has been widely adopted in other scenarios, we would like to mention that it is not suitable to our intrusion detection application, as supervision from the commercial IDSes or hand-crafted rules is very noisy. It is probable for the source of supervision to mislabel a malicious command line as benign. Therefore, even though all the $k$ neighbors in the training set of a test sample have been labeled as benign, it is still probable that the test sample is in fact malicious. In order to tackle this problem, we modify the method to some extent, and compute the average similarity between each test sample $\mathbf t_j$ and its $k$ \emph{malicious} and nearest training neighbors as its intrusion score $o^\mathrm{Retri}_j$. According to experiments, such an innovation leads to obvious performance gains for the retrieval-based method in handling intrusion detection, owing to relief of the negative impact of label noise.

\section{Intrusion Detection Experiments}
\label{sec:exp}

\textbf{Data.}
We collected command-line logs in a production cloud service. We logged the command lines of all the users on $\sim100000$ machines for a week (from May 1$^\mathrm{st}$ to May 7$^\mathrm{th}$, 2022) as the training set, and data from May 29$^\mathrm{th}$ to May 31$^\mathrm{st}$ as the test set. The training and test set contains 30 million and 10 million samples, respectively. The disc space taken by the raw training and test files of command lines is 10.5GB and 4.6GB, respectively.
Bashlex\footnote{https://github.com/idank/bashlex} is utilized to parse the command line data as described in Section~\ref{sec:preprocess}. 
Labels used for supervised fine-tuning came from a commercial IDS, developed by a Fortune Global 500 company and widely deployed in many cloud services. Since there exist many duplicate samples in the test set, and, in order to make the evaluation more effective and more reasonable, we de-duplicate the test set before calculating the concerned metrics to avoid focusing only on common threats in evaluation.

\textbf{Implementation details. } 
The architecture of our command-line language model for experiments is the same as that of BERT-base~\cite{devlin2018bert}, \ie, it involves \num{12} transformer blocks, and in each transformer block there are \num{12} attention heads and the hidden size is \num{768}. 
BPE is used for tokenization and the vocabulary size is set to \num{50000}. 
All command lines that exceed the maximum allowed number of tokens, \ie, \num{1024}, will be trimmed before being fed into the model.
For classification-based tuning, either processing a single command line or temporarily contiguous multiple command lines as model input(s), we need a classification head to process the ``\texttt{[CLS]}'' token embedding obtained from the backbone for final classification. In our experiments, the classification head is a two-layer perceptron initialized by Kaiming's method~\cite{he2015delving}, and it is tuned with a learning rate of 5e-5 for 5 epochs using AdamW, with the language model being frozen.
The multi-line method leverages three temporally contiguous command lines and concatenates them as a single input for model processing.
For reconstruction-based tuning, as has been described in Section~\ref{sec:recon}, we obtained $W$ using SVD each time $f(\cdot)$ has been fine-tuned, and $f(\cdot)$ was fine-tuned by minimizing Eq.~\eqref{eq:loss_recons} once the matrix $W$ was updated. The process was repeated continuously until convergence, and, in general, we found that repeating the process five times suffices. 
We let 95\% of components to be kept by PCA for reconstruction-based tuning, and, for retrieval, we performed $1$NN. That is, the similarity between each test sample and its nearest malicious sample in the training set is adopted as its intrusion score.

\subsection{Quantitative Analyses}
\label{sec:quantitative}

It is of interest to evaluate an IDS from two perspectives: 1) how many intrusions that have not yet been confirmed by the source of supervision, \ie, an existing commercial IDS, can be successfully detected by our IDS; 2) how well does the IDS perform on detecting all intrusions, including those that have already been previously confirmed by the commercial IDS. The former focuses on ``\textbf{out-of-box}'' intrusions while the later further takes ``\textbf{in-box}'' intrusions into consideration. 
Table~\ref{tab:performance} and~\ref{tab:performance-top} compare all methods (introduced in Section~\ref{sec:supervision}) that use supervision. $\mathrm{P_O}@v$ in the table indicates the precision of the top out-of-box predictions of the model. If $v=100$, we calculate the precision of top-100 out-of-box predictions of each method.
In addition to $\mathrm{P_O}@v$, we also evaluate the precision when each method is able to recall $u$ (for $u\approx 100\%$) of all intrusions detected by the commercial IDS. This is achieved by setting a specific intrusion detection threshold for each method according to its prediction scores.
$\mathrm{P_O}$ and $\mathrm{P_{O\&I}}$ calculate the out-of-box precision and overall precision of each method, respectively, under such a threshold.

In Table~\ref{tab:performance}, reconstruction-based tuning, classification-based tuning, and the retrieval-based method all obtain reasonably high $\mathrm{P_{O\&I}}$, when setting a threshold that guarantees almost all in-box intrusions (that have already been found out by the commercial IDS) show higher scores than it. 

For detecting out-of-box intrusions that should have been spotted by the commercial IDS, classification-based model tuned on single command line data shows a final $\mathrm{P_O}@100$ of 100\%, a $\mathrm{P_O}@1000$ of 94.9\%, a $\mathrm{P_O}$ of 83.2\% and a $\mathrm{P_{O\&I}}$ of 99.4\% on average, with some variance over multiple runs. Tuning on multi-line data leads to even more superior performance on the top predictions (with a $\mathrm{P_O}@100$ of 100\% and a $\mathrm{P_O}@1000$ of 99.8\%). The superiority of multi-line classification over single-line classification in their top predictions indicates that in-context information is indeed beneficial. Note that, after de-duplication, the number of test samples is different for the single-line and multi-line test set. On account of this, we do not report the performance in $\mathrm{P_O}$ and $\mathrm{P_{O\&I}}$ for multi-line classification, since it should be unfair and its results should be uncomparable to others.

\begin{table}[t]
    \caption{Quantitative comparison of different methods. Average performance over five runs is reported together with the standard derivation (except for the retrieval-based method which does not require fine-tuning and thus only a single run of test was performed for it).} \vskip -0.1in
    \label{tab:performance} 
    \centering
    \begin{tabular}{lllll} 
    \hline
               & $\mathrm{P_O}$ & $\mathrm{P_{O\&I}}$     \\ 
    \hline
    Reconstruction & $\mathbf{0.913\pm 0.050}$ & $\mathbf{0.999 \pm 0.000}$\\
    Classification & $0.832\pm 0.070$ & $0.994 \pm 0.003$ \\
    Classification (multi)  & - & - \\ 
    Retrieval & 0.569 & 0.892\\
    \hline
    \end{tabular} \vskip -0.1in
\end{table}

\begin{table}[t]
    \caption{Quantitative comparison of different methods on their top-100 and top-1000 out-of-box predictions.} \vskip -0.1in
    \label{tab:performance-top} 
    \centering
    \begin{tabular}{lllll} 
    \hline
               & $\mathrm{P_O}$@100 & $\mathrm{P_O}$@1000    \\ 
    \hline
    Reconstruction & $0.984 \pm 0.032$ & $0.535 \pm 0.092$   \\
    Classification & $\mathbf{1.000 \pm 0.000}$  & $0.949 \pm 0.003$   \\
    Classification (multi)  & $\mathbf{1.000 \pm 0.000}$  & $\mathbf{0.998 \pm 0.001}$ \\
    Retrieval & 0.970 & 0.569   \\
    \hline
    \end{tabular} \vskip -0.15in
\end{table}

\begin{table*}
\caption{In-box-examples vs out-of-box examples.}  \vskip -0.1in
\label{tab:OOB-samples} 
\centering
\begin{tabular}{L{8.1cm}L{8.8cm}} 
\hline
In-box & Out-of-box  \\ 
\hline
\multirow{2}{*}{\texttt{nc -lvnp {\color{blue}*}}}   &     \multirow{2}{*}{\texttt{nc -ulp {\color{blue}*}}}\\
\multirow{2}{*}{\texttt{masscan {\color{blue}*} -p 0-65535 --rate=1000 >> tmp.txt}}   &   \multirow{2}{*}{\texttt{sh /root/masscan.sh {\color{blue}*} -p 0-65535}}          \\
\multirow{2}{*}{\texttt{bash -i >\& {\color{blue}*} 0>\&1}}   &   \multirow{2}{*}{\texttt{java -cp tmp.jar "bash=bash -i >\& {\color{blue}*}"}}          \\
\multirow{2}{*}{\texttt{export https\_proxy="http:{\color{blue}*}"}}   &   \multirow{2}{*}{\texttt{export https\_proxy="socks5:{\color{blue}*}"}}   \\ [6pt]
\texttt{java -jar tmp.jar -C "bash -c \{echo,{\color{blue}*}\} {\color{red}$\backslash$} \{base64,-d\} \{bash,-i\}"}   &   \texttt{python3 tmp.py -p "bash -c \{echo,{\color{blue}*}\} \{base64,-d\} {\color{red}$\backslash$} \{base,-i\}"}   \\ [7pt]

\hline
\end{tabular} \vskip -0.15in 
\end{table*}

The reconstruction-based and retrieval-based methods seem less effective in the sense of achieving high precision on out-of-box predictions, comparing with classification-based tuning. The results show that the two methods fail to generalize to out-of-box samples as well as classification-based tuning, on account of overfitting.
Reconstruction-based tuning shows a $\mathrm{P_O}@1000$ of 53.5\%, while its $\mathrm{P_O}$ is even higher than it.
In general, reconstruction-based tuning leads to very high scores for all in-box intrusions, and there are not many (roughly 200 on the test set, to be more specific) out-of-box results that could show higher or even comparable scores, while on which the prediction precision is also low.
The retrieval-based method outperforms reconstruction-based tuning in $\mathrm{P_O}@1000$, yet shows lower $\mathrm{P_{O\&I}}$. 
In contrast to these two methods, classification-based tuning shows reasonably good performance in all metrics.

\subsection{Comparison with the Commercial IDS}
\label{sec:compare_to_commercial}

In this subsection, we attempt to compare classification-based tuning (which is quantitatively the best among all the developed methods based on our pre-trained command-line language model) to other methods. We mainly compare to the commercial IDS, which is also the source of supervision, since most existing learning-based methods in the literature focus on detecting new users, which is not comparable in our testing environment.

We propose to compare in the sense of an F1-measure which calculates the harmonic mean of precision and recall. The F1-measure only requires fixed final decisions of the model, making it possible to directly compare the decision accuracy of classification-based tuning and that of the commercial IDS.
The evaluation of recall requires some further effort in addition to the evaluation in Section~\ref{sec:quantitative}, as we will need to know the number of all positives.
For handling this, we try to compare the developed classification-based tuning with the commercial IDS on a set of command lines where in-box and out-of-box intrusions highly likely exist, such that the number of all positives can be reasonably estimated.
Specifically, we opt to evaluating on the set of all predicted positive of our method. Such a set also contains almost all predicted positives of the commercial IDS.
We note that comparison on such a set is not entirely fair (as the test samples are chosen from a subset of samples on which classification-based tuning makes confident predictions), yet it is the best we can try to make the comparison possible on a common benchmark.
Comparing with using all test samples, using such a subset leads to over-optimistic recall for both compared methods, given limited resource for test sample labeling. 

As has been shown in Table~\ref{tab:performance}, given the specific threshold, classification-based tuning should show an average overall precision of 99.4\% on the set of its predicted positives.
We also know that its overall recall on this set should be 100\%, since all true positives in the set, either in-box or out-of-box can all be successfully spotted by the method.
Therefore, the F1-measure of such a method is 99.7\%.

Let us assume that the commercial IDS can spot $S$ intrusions on the whole test set, and there are in total $T$ test samples in the predicted positive set of our method and $\mathrm{P_O}=x$.
Considering that the commercial IDS is only capable of detecting in-box intrusions, it misses on all the $xT-xuS$ out-of-box intrusions and we can obtain its recall $97.4\%$ by approximately calculating $uS/(xT+u(1-x)S)$.
As we have assumed that the precision of the commercial IDS is 100\%, it achieves an F1-measure of 98.7\%. Obviously, the classification-based tuning outperforms the commercial IDS in recall and is slightly inferior in precision, and, overall, thus it is superior in the sense of the F1-measure. In comparison with the commercial IDS, our solution leverages AI and machine learning and, in particular, large-scale pre-training on unlabeled data, which endows superior generalization. Benefiting from superior generalization ability to out-of-box cases, out method leads to a higher F1-measure.

Comparing our IDS to other commercial IDSes (in addition to the one that provides supervision) is challenging, since different IDSes define malicious behaviors differently and it makes little sense to compare them in the context of the proposed evaluation metrics. 
Given noisy supervision from the commercial IDS, we have gained considerably superior performance to that of the supervision source, which confirms the effectiveness of our command-line language model pre-trained on massive data, and we will consider other commercial IDSes as the source of supervision for possible comparison in future work.

\subsection{Qualitative Analyses}
\label{sec:qualitative}

\textbf{Generalization. }The right half of Table~\ref{tab:OOB-samples} reports some out-of-box command lines that were failed to be noticed by the commercial IDS but show high intrusion scores using classification-based tuning, while related in-box samples are provided on the left as references. As can be seen, attackers may decode from base64 and execute suspect commands camouflaged under different programming languages, yet the commercial IDS may succeed with \texttt{java} but failed with \texttt{python3}. This case shows the effectiveness of our method in generalizing across interpreters of different programming languages. 
From an example of the command \texttt{nc}, we can also observe the generalization ability of the proposed method across command flags. It avoids the failure case of ``\texttt{nc -ulp {\color{blue}*}}'' which is functionally very similar to ``\texttt{nc -lvnp {\color{blue}*}}'' and should have been classified into the same class (as malicious or benign). The ``\texttt{export} ...'' example shows how generalization in arguments could help.

\textbf{Preference of different methods. }By comparing the true positives of developed methods, we found that they are sensitive to different out-of-box intrusions. Classification using single command lines as inputs tends to first capture intrusions whose purpose is to bind and reverse shells, \eg, the command lines in Table~\ref{tab:OOB-samples} with ``\texttt{nc}'', while multi-line classification, as has been discussed, is capable of detecting malicious behaviors involving a sequence of suspect command-line operations. Moreover, higher scores should be assigned to command-line sequences involving multiple sensitive operations than that involving only one suspect operation.
Reconstruction-based tuning performs better in detecting intrusions that decode from base64 and execute suspect commands, \eg, ``\texttt{echo {\color{blue}*} | bash64 -d | bash -i}'', where the symbol ``\texttt{{\color{blue}*}}'' indicates a \texttt{bash64}-encoded operation which is not shown for security and privacy reasons. We conjecture that this is due to the fact that \texttt{bash64}-encoded operations are difficult to be reconstructed. As can be seen, these methods complement each other, and an ensemble of all these methods can further boost the out-of-box intrusion detection performance, which should be explored in future work.

\section{Related Work}\label{sec:related}

\textbf{Learning-based intrusion detection. } 
AI-based intrusion detection methods in the literature are mainly designed for small data and focus mostly on detecting new users of the command-line interface.
For instance, Lane and Brodley~\cite{lane1997application} proposed to build a profile that enumerates command names and flags in historical operations for each user and evaluated the similarity of a command operation to all profiles in order to determine whether it is abnormal or not. Huang \etal~\cite{huang2011masquerade} further advocated to align all sequence of commands before creating the profile of each user adopting hidden Markov models. More recently, Liu and Mao~\cite{liu2022new} constructed a sequence-to-sequence model on the basis of recurrent neural networks to predict following command-line behaviors given previous ones, and issued the uncertainty and fluctuation of user behavior by comparing the difference in user profiles.
These methods require abundant data for each possible user and are difficult to quickly adapt to new benign users which, however, widely exist in cloud environments. Moreover, these methods only took partial information of each command line into account. Specifically, Lane and Brodley's and Liu and Mao's only utilize command names and flags while Huang \etal's only utilizes command names.

\textbf{Language models. } 
Often trained in a self-supervised manner on a huge amount of data, language models are becoming increasingly popular. 
Representatives of these models like BERT~\cite{devlin2018bert}, RoBERTa~\cite{liu2019roberta}, GPT~\cite{brown2020language}, and CLIP~\cite{radford2021learning} have dominated certain applications. 
Taking BERT as an example, it has been widely used in English-based query on the Google search engine~\cite{search2020}, and it has renovated the pipeline of model training in a variety of NLP applications. 
Despite the remarkable successes, there is little work that thoroughly discusses self-supervised learning and language models for intrusion detection and cloud security. This work encourages to marry language models to IDSes and compares different intrusion detection methods based on a command-line language model.
Somewhat related, Setianto and Tsani~\cite{setianto2021gpt} proposed to fine-tune a pretrained GPT model on a question answering task, where the context is shell commands and the answer is the Unix utility to automatically parse command lines from
honeypots, which is possible to be incorporated into future IDSes. It shows another potential way of using language model for IDSes, although it does not target at improving the intrusion detection precision and recall directly.

\section{Conclusions}
In this paper, we aim at adopting AI to intrusion detection at scale. 
Equipped with an advanced command-line language model model, several different methods have been proposed for achieving remarkable in-box and out-of-box intrusion detection performance. 
The effectiveness of our IDS is verified on tens of millions of command lines collected in a production cloud environment. Under the premise of successfully recalling nearly all in-box intrusions, our method is capable of detecting a reasonably large number of out-of-box intrusions with a favorable precision.
Qualitative analyses have also been performed to shed light on the reason of the effectiveness of our method.  
\bibliographystyle{plain}
\bibliography{ref}

\begin{thebibliography}{10}

\bibitem{bottou1998online}
L{\'e}on Bottou et~al.
\newblock Online learning and stochastic approximations.
\newblock {\em On-line learning in neural networks}, 17(9):142, 1998.

\bibitem{brown2020language}
Tom Brown, Benjamin Mann, Nick Ryder, Melanie Subbiah, Jared~D Kaplan, Prafulla
  Dhariwal, Arvind Neelakantan, Pranav Shyam, Girish Sastry, Amanda Askell,
  et~al.
\newblock Language models are few-shot learners.
\newblock {\em Advances in neural information processing systems},
  33:1877--1901, 2020.

\bibitem{christodorescu2005semantics}
Mihai Christodorescu, Somesh Jha, Sanjit~A Seshia, Dawn Song, and Randal~E
  Bryant.
\newblock Semantics-aware malware detection.
\newblock In {\em 2005 IEEE symposium on security and privacy (S\&P'05)}, pages
  32--46. IEEE, 2005.

\bibitem{cover1967nearest}
Thomas Cover and Peter Hart.
\newblock Nearest neighbor pattern classification.
\newblock {\em IEEE transactions on information theory}, 13(1):21--27, 1967.

\bibitem{devlin2018bert}
Jacob Devlin, Ming-Wei Chang, Kenton Lee, and Kristina Toutanova.
\newblock Bert: Pre-training of deep bidirectional transformers for language
  understanding.
\newblock {\em arXiv preprint arXiv:1810.04805}, 2018.

\bibitem{he2015delving}
Kaiming He, Xiangyu Zhang, Shaoqing Ren, and Jian Sun.
\newblock Delving deep into rectifiers: Surpassing human-level performance on
  imagenet classification.
\newblock In {\em Proceedings of the IEEE international conference on computer
  vision}, pages 1026--1034, 2015.

\bibitem{huang2011masquerade}
Lin Huang and Mark Stamp.
\newblock Masquerade detection using profile hidden markov models.
\newblock {\em computers \& security}, 30(8):732--747, 2011.

\bibitem{lane1997application}
Terran Lane and Carla~E Brodley.
\newblock An application of machine learning to anomaly detection.
\newblock In {\em Proceedings of the 20th national information systems security
  conference}, volume 377, pages 366--380. Baltimore, USA, 1997.

\bibitem{lecun1988theoretical}
Yann LeCun, D~Touresky, G~Hinton, and T~Sejnowski.
\newblock A theoretical framework for back-propagation.
\newblock In {\em Proceedings of the 1988 connectionist models summer school},
  volume~1, pages 21--28, 1988.

\bibitem{li2020sentence}
Bohan Li, Hao Zhou, Junxian He, Mingxuan Wang, Yiming Yang, and Lei Li.
\newblock On the sentence embeddings from pre-trained language models.
\newblock {\em arXiv preprint arXiv:2011.05864}, 2020.

\bibitem{liu2008isolation}
Fei~Tony Liu, Kai~Ming Ting, and Zhi-Hua Zhou.
\newblock Isolation forest.
\newblock In {\em 2008 eighth ieee international conference on data mining},
  pages 413--422. IEEE, 2008.

\bibitem{liu2022new}
Wei Liu, Yu~Mao, Linlin Ci, and Fuquan Zhang.
\newblock A new approach of intrusion detection with command
  sequence-to-sequence model.
\newblock In {\em Advances in Intelligent Data Analysis and Applications},
  pages 169--182. Springer, 2022.

\bibitem{liu2019roberta}
Yinhan Liu, Myle Ott, Naman Goyal, Jingfei Du, Mandar Joshi, Danqi Chen, Omer
  Levy, Mike Lewis, Luke Zettlemoyer, and Veselin Stoyanov.
\newblock Roberta: A robustly optimized bert pretraining approach.
\newblock {\em arXiv preprint arXiv:1907.11692}, 2019.

\bibitem{natarajan2013learning}
Nagarajan Natarajan, Inderjit~S Dhillon, Pradeep~K Ravikumar, and Ambuj Tewari.
\newblock Learning with noisy labels.
\newblock In {\em Advances in neural information processing systems}, 2013.

\bibitem{radford2021learning}
Alec Radford, Jong~Wook Kim, Chris Hallacy, Aditya Ramesh, Gabriel Goh,
  Sandhini Agarwal, Girish Sastry, Amanda Askell, Pamela Mishkin, Jack Clark,
  et~al.
\newblock Learning transferable visual models from natural language
  supervision.
\newblock In {\em International Conference on Machine Learning}, pages
  8748--8763. PMLR, 2021.

\bibitem{reimers2019sentence}
Nils Reimers and Iryna Gurevych.
\newblock Sentence-bert: Sentence embeddings using siamese bert-networks.
\newblock {\em arXiv preprint arXiv:1908.10084}, 2019.

\bibitem{rumelhart1985learning}
David~E Rumelhart, Geoffrey~E Hinton, and Ronald~J Williams.
\newblock Learning internal representations by error propagation.
\newblock Technical report, California Univ San Diego La Jolla Inst for
  Cognitive Science, 1985.

\bibitem{scholkopf2001estimating}
Bernhard Sch{\"o}lkopf, John~C Platt, John Shawe-Taylor, Alex~J Smola, and
  Robert~C Williamson.
\newblock Estimating the support of a high-dimensional distribution.
\newblock {\em Neural computation}, 13(7):1443--1471, 2001.

\bibitem{search2020}
Barry Schwartz.
\newblock Google: Bert now used on almost every english query, [Accessed
  August, 2022].
\newblock
  \url{https://searchengineland.com/google-bert-used-on-almost-every-english-query-342193}.

\bibitem{sennrich2015neural}
Rico Sennrich, Barry Haddow, and Alexandra Birch.
\newblock Neural machine translation of rare words with subword units.
\newblock {\em arXiv preprint arXiv:1508.07909}, 2015.

\bibitem{setianto2021gpt}
Febrian Setianto, Erion Tsani, Fatima Sadiq, Georgios Domalis, Dimitris
  Tsakalidis, and Panos Kostakos.
\newblock Gpt-2c: a parser for honeypot logs using large pre-trained language
  models.
\newblock In {\em Proceedings of the 2021 IEEE/ACM International Conference on
  Advances in Social Networks Analysis and Mining}, pages 649--653, 2021.

\bibitem{tenney2019bert}
Ian Tenney, Dipanjan Das, and Ellie Pavlick.
\newblock Bert rediscovers the classical nlp pipeline.
\newblock {\em arXiv preprint arXiv:1905.05950}, 2019.

\bibitem{vaswani2017attention}
Ashish Vaswani, Noam Shazeer, Niki Parmar, Jakob Uszkoreit, Llion Jones,
  Aidan~N Gomez, {\L}ukasz Kaiser, and Illia Polosukhin.
\newblock Attention is all you need.
\newblock {\em Advances in neural information processing systems}, 30, 2017.

\bibitem{wall2003singular}
Michael~E Wall, Andreas Rechtsteiner, and Luis~M Rocha.
\newblock Singular value decomposition and principal component analysis.
\newblock In {\em A practical approach to microarray data analysis}, pages
  91--109. Springer, 2003.

\bibitem{ye2017survey}
Yanfang Ye, Tao Li, Donald Adjeroh, and S~Sitharama Iyengar.
\newblock A survey on malware detection using data mining techniques.
\newblock {\em ACM Computing Surveys (CSUR)}, 50(3):1--40, 2017.

\end{thebibliography}

\end{document}